\documentclass[aps,pra,twocolumn,floatfix,longbibliography]{revtex4-2}

\usepackage{amsmath}
\usepackage{amssymb}
\usepackage{graphicx}
\graphicspath{{pictures/}{v6/pictures/}{./pictures/}{./v6/pictures/}{}}
\usepackage{hyperref}
\usepackage{booktabs}
\usepackage{multirow}
\usepackage{natbib}
\usepackage[english]{babel}

\hypersetup{
    colorlinks=true,
    linkcolor=blue,
    citecolor=blue,
    urlcolor=blue
}

\begin{document}
\title{Nonthermal Dynamics of Rydberg Atom Chains with Constrained Four-Body Interactions}
\author{Tianyi Yan}
\author{Weibin Li}
\affiliation{School of Physics and Astronomy and Centre for the Mathematics and Theoretical Physics of Quantum Non-equilibrium Systems, University of Nottingham, Nottingham NG7 2RD, United Kingdom}

\begin{abstract}
We investigate the dynamics of a linear chain of Rydberg atoms driven by a constrained four-body interaction, where two neighboring atoms are excited simultaneously from the electronic ground state $|0\rangle$ to Rydberg state $|1\rangle$ only when their closest neighbors are in $|0\rangle$ state. By employing an ansatz for the many-body ground state, the many-body ground state energy, which scales linearly with the chain length $L$, is obtained analytically in the thermodynamic limit and agrees with that of exact diagonalization. Our model supports quantum many-body scar eigenstates with nearly equally spaced energies. The scar states overlap strongly with the basis state $|\mathbf{0}\rangle=|0\cdots 0\rangle$. We show that the overlap distribution is tilted by the state-dependent four-body interaction, leading to nonergodic dynamics. The four-body constrained model can be realized with Rydberg atoms in a Peierls array with alternating bond lengths, where atoms on the shorter and longer bonds experience Rydberg blockade and antiblockade, respectively. Our study provides a pathway to investigate constrained nonergodic dynamics with four-body interactions by exploiting the programmable Rydberg atom arrays.
\end{abstract}

\maketitle

\section{Introduction}
\label{sec:intro}

Many-body systems with multi-body interactions often display exotic spectra and serve as excellent examples for studying multipartite entanglement~\cite{peng_ground-state_2010, giampaolo_genuine_2014} and quantum phase transitions~\cite{pachos_three-spin_2004, peng_quantum_2009, zhang_three-body_2011, you_emergent_2017, xu_multipartite_2024}. Such interactions have been theoretically engineered using Floquet methods~\cite{verga_entanglement_2023, petiziol_quantum_2021} and experimentally demonstrated with trapped ions recently~\cite{katz_n_2022, katz_demonstration_2023, katz_programmable_2023}. They have also proven valuable for simulating lattice gauge theories on cold atom platforms~\cite{dai_four-body_2017, yang_observation_2020, zhou_thermalization_2022}. Recently, Rydberg platforms have emerged as an ideal system for engineering multi-body interactions by leveraging the long-range dipole interactions between Rydberg atoms. A celebrated paradigm is the PXP model~\cite{moessner_ising_2001, fendley_competing_2004, lesanovsky_many-body_2011, khemani_signatures_2019, choi_emergent_2019, lin_slow_2020, verresen_prediction_2021, karle_area-law_2021, surace_exact_2021, su_observation_2023, ivanov_volume-entangled_2025} that features a three-body interaction, which has been realized using strongly interacting Rydberg atoms~\cite{bernien_probing_2017, bluvstein_controlling_2021}. In Rydberg systems, the strong interaction-induced blockade imposes kinetic constraints~\cite{saffman_quantum_2010, shao_rydberg_2024}, resulting in an exotic spectrum that weakly violates ergodicity~\cite{palmer_models_1984, fredrickson_kinetic_1984} and the eigenstate thermalization hypothesis (ETH)~\cite{deutsch_quantum_1991, srednicki_chaos_1994, rigol_thermalization_2008, polkovnikov_colloquium_2011, dalessio_quantum_2016}. Consequently, certain initial states exhibit nonergodic dynamics arising from high overlaps with quantum many-body scars (QMBS)~\cite{turner_weak_2018, turner_quantum_2018, lin_exact_2019, serbyn_quantum_2021, yao_quantum_2022, moudgalya_quantum_2022, chandran_quantum_2023}. Recent studies have expanded nonergodic dynamics beyond three-body interactions~\cite{kerschbaumer_quantum_2025, hosseinabadi_kinetically_2026}. Despite the growing interest in engineering multi-body interactions on Rydberg platforms, most studies rely on the blockade mechanism. In contrast, the concept of the Rydberg antiblockade mechanism has been well established~\cite{ates_antiblockade_2007, amthor_evidence_2010, li_nonadiabatic_2013, marcuzzi_facilitation_2017, su_rydberg_2020, su_dipole-dipole-interactiondriven_2021, ding_facilitation-induced_2023, zhang_quantum_2024}. Its use for realizing experimentally accessible multi-body models---particularly those exhibiting nonergodic dynamics---remains largely unexplored~\cite{datla_statistical_2026}.

\begin{figure}[thbp]
    \centering
    \includegraphics[width=0.95\linewidth]{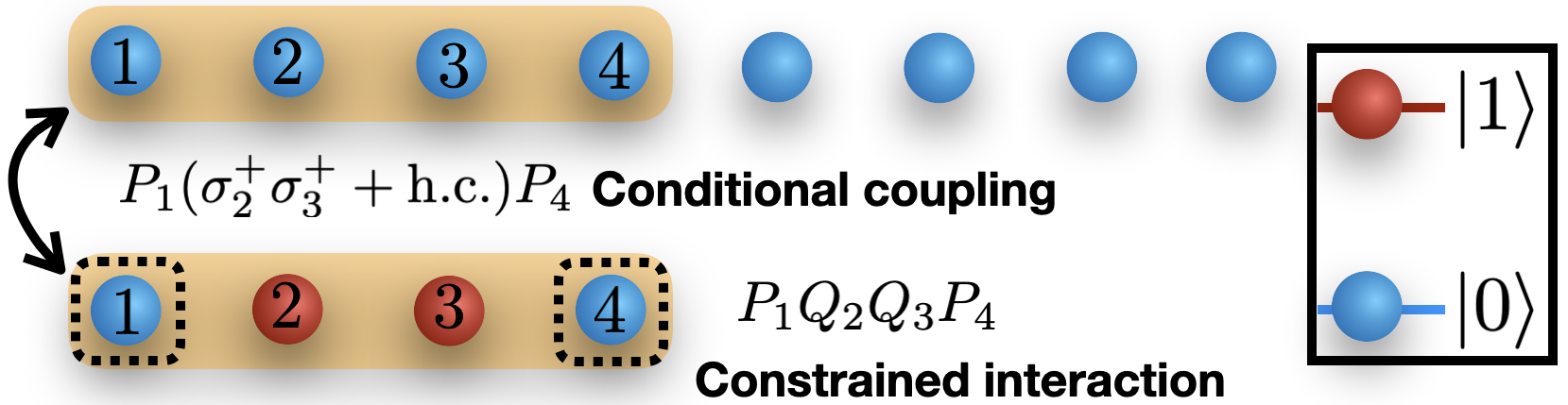}
    \caption{Constrained four-body interaction on a one-dimensional chain of spin-$1/2$ atoms. In a block of four sites, i.e. a tetramer, the central pair can flip their spin state only if the two boundary atoms are in $|0\rangle$ state. After the pair is excited to state $|1\rangle$ simultaneously the tetramer acquires a configuration-dependent FB interaction. }
    \label{fig1}
\end{figure}

In this work, we investigate the many-body ground state and weakly ergodicity-broken dynamics in a spin-half chain (length $L$) with kinetically constrained four-body (FB) interactions. In our model, a pair of neighboring atoms is simultaneously excited from state $|0\rangle$ to $|1\rangle$ only when their closest neighbors are both in the $|0\rangle$ state, giving rise to the kinetically constrained excitation. These four atoms (forming a tetramer) then experience a configuration-dependent FB interaction after the excitation of the pair. Both constrained processes of the tetramers are shown in Fig. ~\ref{fig1}. We develop an ansatz for the many-body ground state and find that the ground state energy grows linearly with the system size $L$ as $L\to \infty$. The entanglement entropy shows periodic high and low values with subsystem size, capturing the strongly entangled bonds within the central pair of a tetramer. We show that the  constrained FB interaction in the Hamiltonian modifies the spectrum of the kinetic term, producing a strip of eigenstates with high overlap with many-body product state $|\mathbf{0}\rangle=|0\cdots 0\rangle$. Starting from this initial state, nonthermal dynamics are evidenced by the periodic revival of the Loschmidt echo. We propose to realize this model using a non-equidistant chain of Rydberg atoms with alternating bond lengths $a$ and $b$ ($a<b$), yielding strong $(V_a)$ and weak $(V_b)$ nearest-neighbor van der Waals interactions. Under the antiblockade condition on the weak bonds ($V_b=2\Delta$), we derive the constrained four-body effective model that exhibits nonergodic dynamics from the fully polarized initial state $|\mathbf{0}\rangle$.  Our study opens a route to studying the multi-body interaction models that show nonergodic dynamics via the Rydberg antiblockade mechanism and is readily implementable on current Rydberg atom array platforms.

\section{Energy and strong pair bound in the many-body ground state}
\label{sec:model}
\label{sec:gs}
Constrained dynamics occur in the tetramers of neighboring atoms, as illustrated in Fig.~\ref{fig1}. Here the atom pair sitting in the center of the tetramer flips its states simultaneously only if the pair has identical spin states and the flanking sites are in state $|0\rangle$. Once the pair is excited to state $|1\rangle$, the tetramer experiences a clustered FB interaction. In a chain with $L$ sites, the dynamics of the atoms is given by the kinetically constrained and interacting Hamiltonian,
\begin{equation}
    \begin{split}
        H= g\sum_{j=1}^{L/2-1}&\left[P_{2j-1}(\sigma_{2j}^{+}\sigma_{2j+1}^{+}+\text{h.c.})P_{2j+2}\right.\\
        &\left. - P_{2j-1}Q_{2j}Q_{2j+1}P_{2j+2}\right],
    \end{split}
    \label{eff_ham}
\end{equation}
where $P_j=|0\rangle_j\langle0|$, $Q_j=|1\rangle_j\langle1|$, $\sigma_j^+=|1\rangle_j\langle0|$ and $\sigma_j^-=|0\rangle_j\langle1|$ on the $j$-th site. The parameter $g$ characterizes the strength of the FB interaction. The first term of the Hamiltonian describes the constrained coupling between the configurations in a tetramer, whereas the second term describes the state-dependent FB interaction.  The realization of this model is discussed towards the end of this article.  

The FB Hamiltonian belongs to a large family of many-body models that are driven by multi-body, cluster interactions~\cite{peng_quantum_2009,leibTransmonQuantumAnnealer2016,chancellorCircuitDesignMultibody2017a,liuSynthesizingThreebodyInteraction2020b,gambettaEngineeringNonBinaryRydberg2020,gambettaLongRangeMultibodyInteractions2020,katz_n_2022,menkeDemonstrationTunableThreeBody2022,katz_demonstration_2023,katz_programmable_2023,andradeEngineeringEffectiveThreespin2022a,luoRealizationThreeFourbody2025,Katz26}. These models often involve competing multi-body interactions with single- or two-body interactions. Our model introduces a kinetically constrained FB coupling that interacts with state-dependent FB interactions. This unique interaction leads to the specific properties of the Hamiltonian. For example, it preserves the total magnetization of the inner pairs within non-consecutive tetramers, i.e., $[H,S]=0$, with the total magnetization of the inner pairs $S=\sum_{j=1,3,5,7...}\sigma^z_{2j}\sigma^z_{2j+1}$. 

\begin{figure}[thbp]
    \centering
    \includegraphics[width=1.0\linewidth]{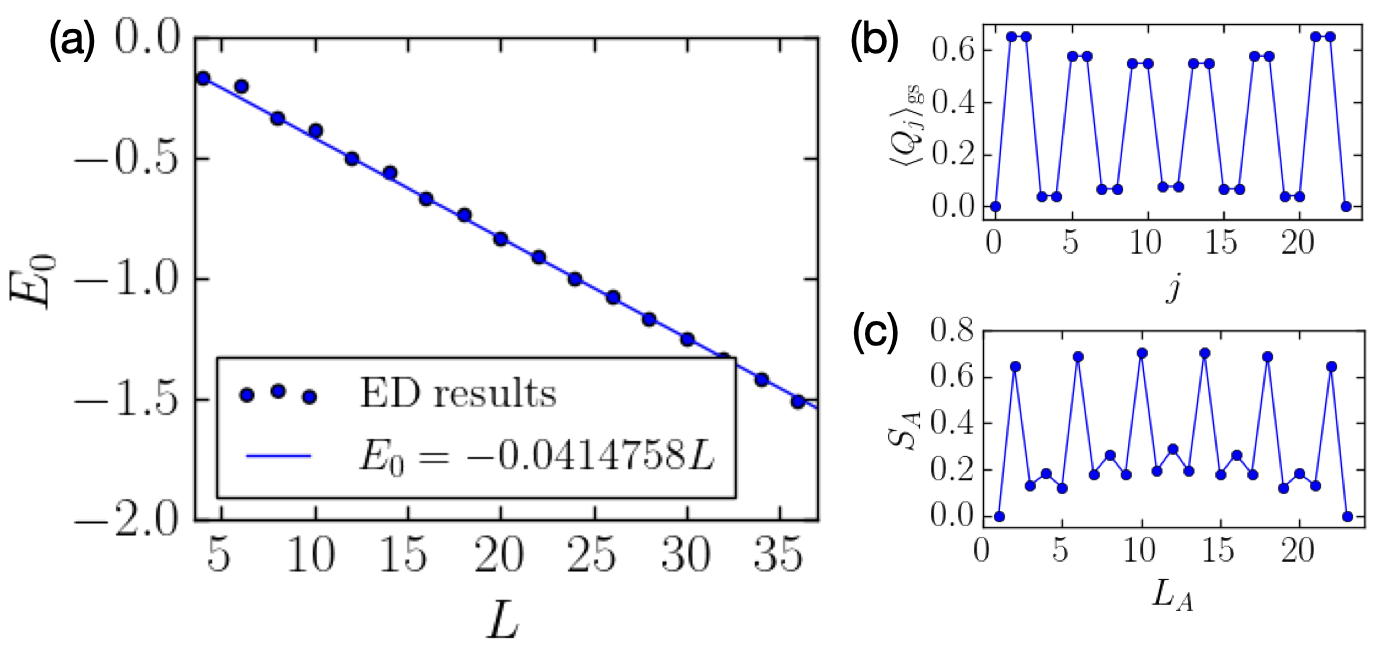}
    \caption{(a) Ground state energy versus system size $L$.  The blue line is the thermodynamic-limit prediction $\bar{E}_0$. (b) Site-resolved Rydberg occupation of the ground state versus site index. Due to the attractive FB interaction and open boundary condition, central pairs of the tetramers have higher excitation probabilities. (c) Bipartite entanglement entropy of the ground state versus subsystem size. The relatively higher entanglement is found between inner pairs of the tetramer. In all panels, we consider $g=0.1$.}
    \label{fig2}
\end{figure}

We develop an exact wavefunction
ansatz~\cite{iadecola_quantum_2019,ovchinnikov_antiferromagnetic_2003} of the ground state by introducing
the orthonormal basis states,
\begin{equation}
    |\psi_m\rangle = \frac{1}{\sqrt{w_m}}\bigl(\mathcal{P}\Sigma^{+}\bigr)^m|{\bf 0}\rangle,
    \quad m=0,2,4,\ldots,
    \label{eq:ansatz}
\end{equation}
where $w_m=[(L/2-m/2-1)!\,(L/2)]/[(m/2)!\,(L/2-m)!]$ is a normalization factor,
$\mathcal{P}=\sum_{j=1}^{L/2-1}P_{2j-1}P_{2j+2}$, and
$\Sigma^{+}=\sum_{j=1}^{L/2-1}\sigma_{2j}^{+}\sigma_{2j+1}^{+}$. In the basis
$\{|\psi_m\rangle\}$, $H$ takes a tridiagonal form with
diagonal elements,
\begin{equation}
    \langle\psi_m|H|\psi_m\rangle = -\tfrac{g}{2}\,m \nonumber,
    \label{eq:diag}
\end{equation}
(an overall constant $gL/2$ is not considered explicitly in  the energy) and off-diagonal elements,
\begin{equation}
    \langle\psi_{m'}|H|\psi_m\rangle = g\bigl(s_m\,\delta_{m',m-2}
    +s_{m+2}\,\delta_{m',m+2}\bigr), \nonumber
    \label{eq:offdiag}
\end{equation}
where $s_m=\sqrt{m(L/2-m+2)(L/2-m+1)/(L-m)}$. For such tridiagonal matrix~\cite{hollenberg_analytic_1996}, we obtain the ground state energy,
\begin{equation}
    \lim_{L\to \infty} E_0 = -g\left(\frac{m}{2} + 2\,s_m\right).
    \label{eq:E0}
\end{equation}
In the thermodynamic limit $L\gg 1$, the infimum is attained at $m^*\approx 0.2689\,L$
with $s_{m^*}\approx 0.1402\,L$, yielding a ground state energy that changes linearly with
$L$, $\bar{E}_0 \approx -0.4148\,g\,L.$ We compare the analytical result with exact diagonalization calculation. As shown in Fig.~\ref{fig2}(a), the ground state energy gradually approaches $\bar{E}_0 $ when increasing $L$. When $L>24$, this linear scaling is in excellent agreement with exact diagonalization. 

The populations in the chain are strongly affected by the constrained interaction. As an example, we show the distribution $\langle Q_j\rangle$ with chain length $L=24$  in Fig.~\ref{fig2}(b). Starting from the left (right) boundary, every four sites form a tetramer. In each tetramer, the inner pair occupies a high population in state $|1\rangle$ with $\langle Q_j\rangle>0.5$, while the atoms at the two ends remain weakly excited $\langle Q_j\rangle<0.1$, driven by the fact that such configuration lowers the overall energy. At the edge of the chain, the atoms are strictly in state $|0\rangle$. Here any occupation in state $|1\rangle$ vanishes the projector involving $P_1$ and $P_L$, which increases the energy. The zero edge population (i.e. $\langle Q_1\rangle = \langle Q_L\rangle=0$) is a direct consequence of the constrained interaction. 

As the chain contains $K$ tetramers, each pair in the middle of a tetramer is strongly bound. We characterize the bound with the bipartite entanglement entropy of the ground state $S_A$ between subsystem  $1\le L_A<L$ (from left to right)  and $L_B=L-L_A$, as shown in Fig.~\ref{fig2}(c), respectively.  The entanglement entropy displays a characteristic alternating pattern with a period $4$. When the right (left) boundary of subchain $A$ ($B$) is on the edge of a tetramer, the entanglement is relatively low. Higher  entanglement, at $L_A=2, 6, 10 \cdots$, is found when we  break an inner pair of atoms in a tetramer, reflecting the strong entanglement inside the tetramer.

\begin{figure}[thbp]
    \centering
    \includegraphics[width=1.0\linewidth]{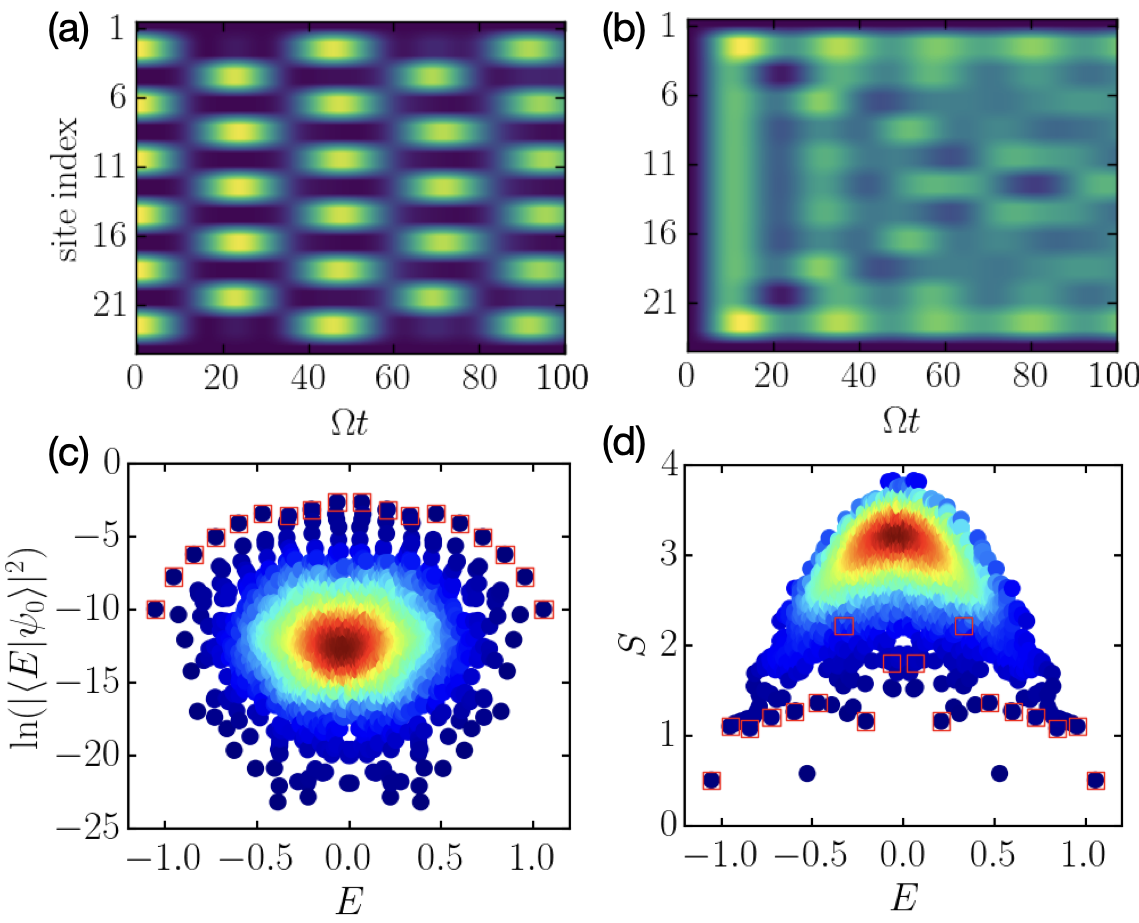}
    \caption{Dynamics and spectrum of Hamiltonian $H_{\text{kin}}$. (a) Occupation $\langle Q_i\rangle$ starting from $|\mathbb{Z}_4\rangle$. (b) Occupation $\langle Q_i\rangle$ starting from $|\mathbf{0}\rangle$. (c) Overlap of $|\mathbb{Z}_4\rangle$ with the eigenstates. (d) Bipartite entanglement entropy of the eigenstates. In all panels, $g=0.1$. }  \label{fig3}
\end{figure}
\section{Nonthermal dynamics of the kinetically constrained FB model}
\label{sec:dynamics}

Given the highly constrained structure of $H$, nonthermal dynamics is expected. Without the FB interaction, the constrained coupling (i.e. the first term on the right-hand side of Eq.~(\ref{eff_ham})) is analogous to the PXP model. To distinguish the roles played by the two terms in Eq.~(\ref{eff_ham}), we first analyze the dynamics by considering the kinetically constrained flip term alone, and then restore the state-dependent FB interaction. 
\begin{figure}[thbp]
    \centering
    \includegraphics[width=1\linewidth]{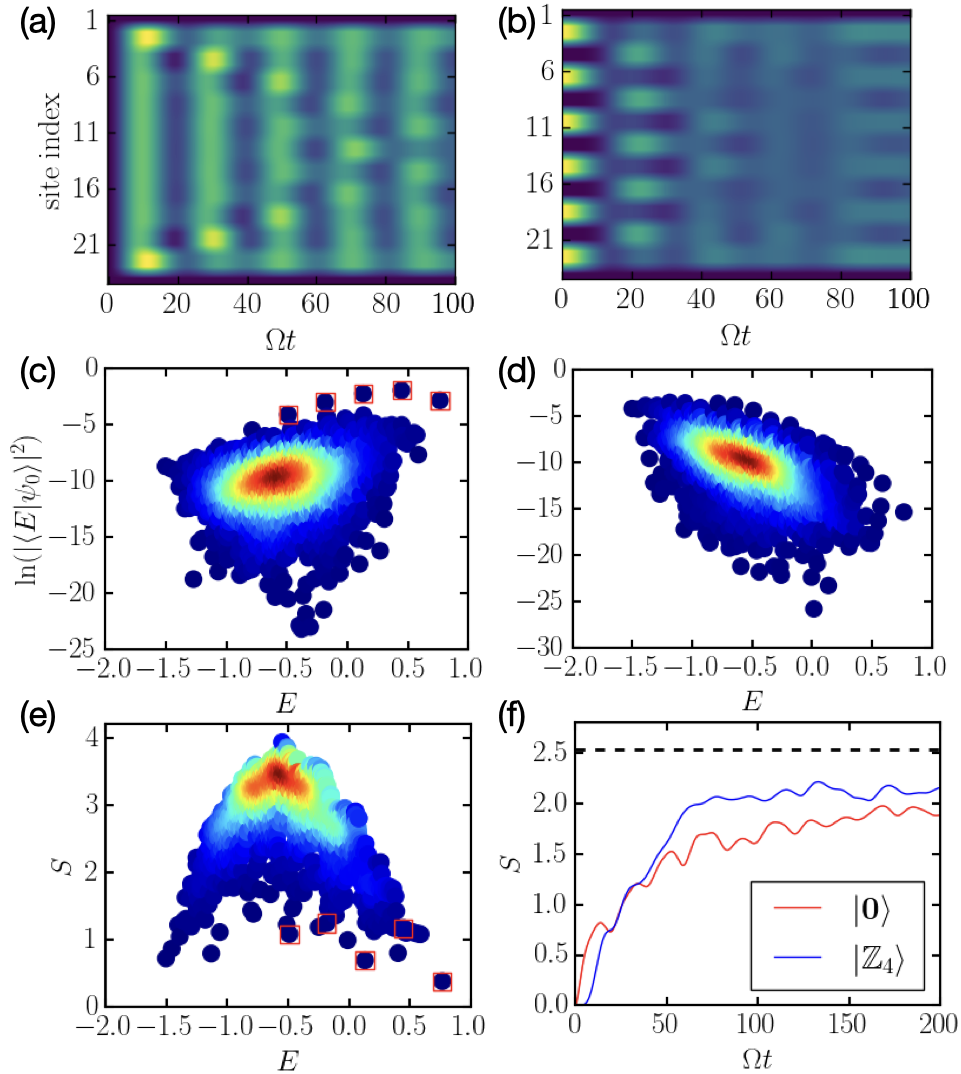}
    \caption{Dynamics and spectrum of the full model, Eq.~(\ref{eff_ham}). (a) Occupation $\langle Q_i\rangle$ starting from state $|\mathbf{0}\rangle$. (b) Occupation $\langle Q_i\rangle$ starting from state $|\mathbb{Z}_4\rangle$. (c) Overlap between states $|\mathbf{0}\rangle$ and the eigenstates. (d) Overlap of $|\mathbb{Z}_4\rangle$ with the eigenstates. (e) Bipartite entanglement entropy of the eigenstates; the dashed line indicates the Page value $\approx 2.54$. (f) Entanglement-entropy dynamics. Panels (a) and (b) use $L=24$. Panels (c)--(f) use $L=36$. In all panels, $g=0.1$.}
    \label{fig4}
\end{figure}

\subsection{Symmetric quantum many-body scar}
\label{sec:dyn-kin}
\begin{figure}[thbp]
    \centering
    \includegraphics[width=1\linewidth]{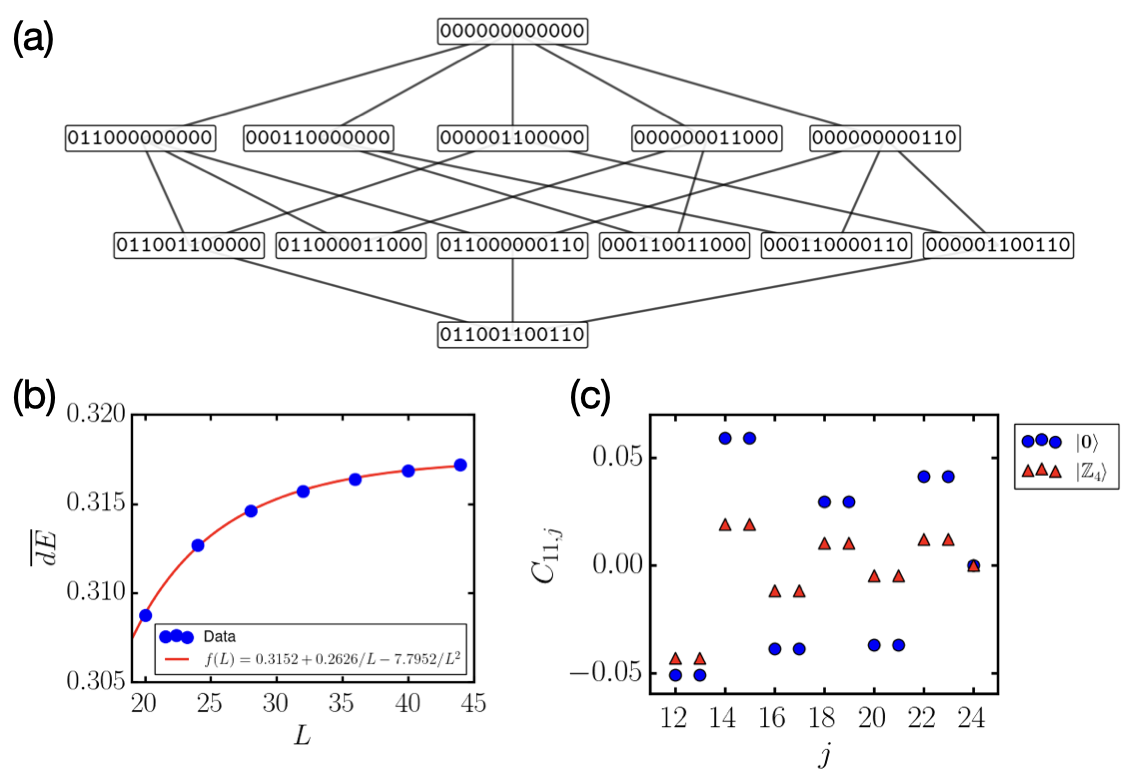}
    \caption{Scar structure of Eq.~(\ref{eff_ham}). (a) Connectivity graph in the allowed computational basis. (b) Average energy spacing among the four eigenstates with the largest overlap on $|\mathbf{0}\rangle$, versus chain length $L$. (c) Connected correlations $C_{i,j}=\langle Q_i Q_j\rangle-\langle Q_i\rangle\langle Q_j\rangle$ between site $11$ and site $j$ for $|\mathbf{0}\rangle$ and $|\mathbb{Z}_4\rangle$, taken at $g t=35$ and $30$, respectively, for $L=24$. In (b) and (c), $g=0.1$.}
    \label{fig5}
\end{figure}
\begin{figure*}[t]
    \centering
    \includegraphics[width=0.95\textwidth]{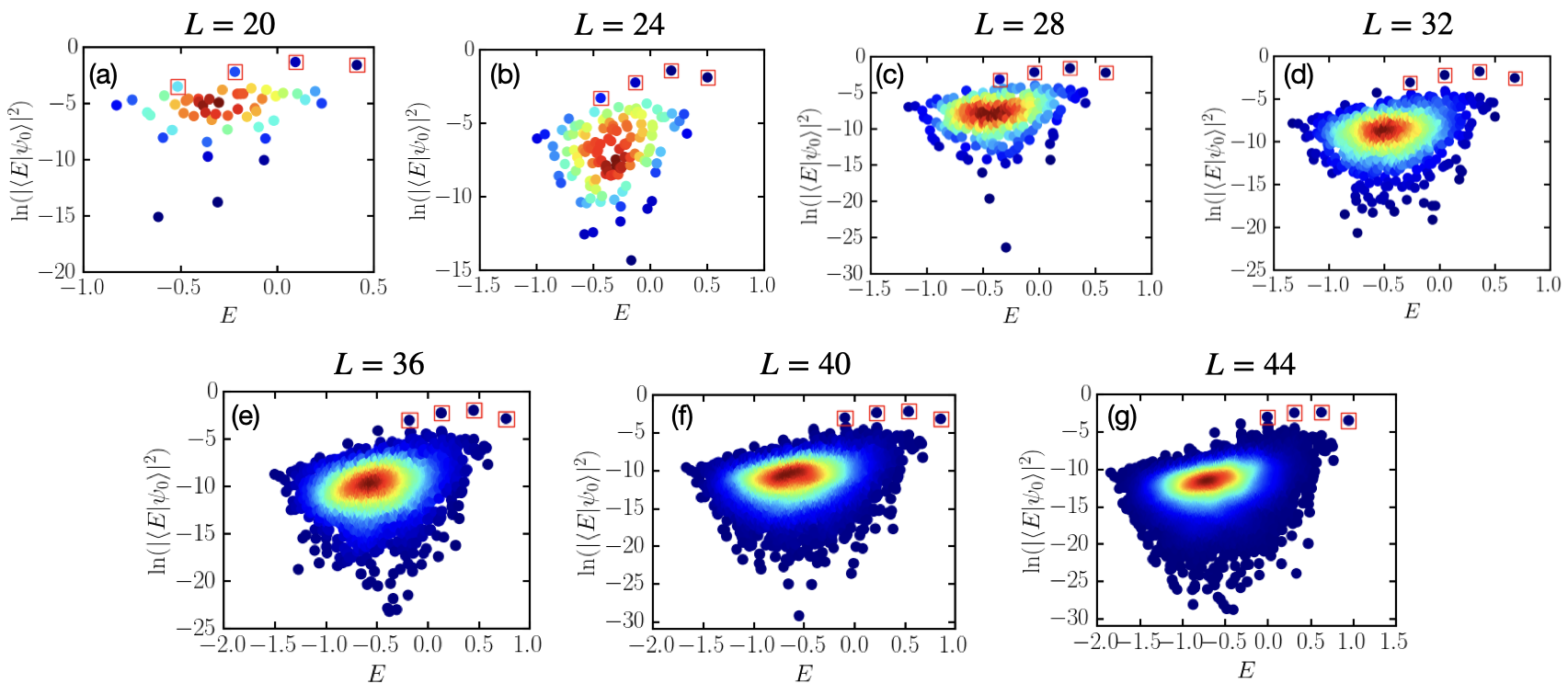}
    \caption{Overlaps $\ln|\langle E|\mathbf{0}\rangle|^2$ of state $|\mathbf{0}\rangle$ with the eigenstates of Eq.~(\ref{eff_ham}). In (a)--(g) chain lengths $L=20$ to $L=44$ in steps of $4$. Red squares mark the four highest-overlap eigenstates, which remain nearly equally spaced with $\Delta E\approx0.3$ independent of $L$. }
    \label{fig6}
\end{figure*}
Omitting the constrained FB interaction in Eq.~(\ref{eff_ham}) leaves the kinetically constrained  Hamiltonian describing the spin flipping process,
\begin{equation}
    H_{\text{kin}} = g\sum_{j=1}^{L/2-1}P_{2j-1}(\sigma_{2j}^{+}\sigma_{2j+1}^{+}+\text{h.c.})P_{2j+2},
    \label{non_int_eff_ham}
\end{equation}
which resembles a clustered PXP model. In the dynamics, this model exhibits persistent oscillations for the $|\mathbb{Z}_4\rangle = |011001100\dots\rangle$ state. As shown in Fig.~\ref{fig3}(a), the central pairs within each tetramer display clear revivals, reminiscent of the $|\mathbb{Z}_2\rangle$ oscillations. In contrast, the polarized state $|\mathbf{0}\rangle$ does not exhibit coherent revivals [Fig.~\ref{fig3}(b)].

The origin of these revivals can be observed in the spectrum. Fig.~\ref{fig3}(c) shows the overlaps between $|\mathbb{Z}_4\rangle$ and the eigenstates of $H_{\text{kin}}$: a strip of high overlaps is spaced approximately $0.1$ apart in energy. They are symmetric with respect to $E=0$. Most of these eigenstates have low entanglement entropies [Fig.~\ref{fig3}(d)], as expected for quantum many-body scars, and they are responsible for the nonthermal dynamics of state $|\mathbb{Z}_4\rangle$. $H_{\text{kin}}$ possesses a chiral symmetry $\mathcal{C} = \prod_{j=1}^{L/2-1} \sigma_{2j}^z$, which anticommutes with the Hamiltonian and yields a spectrum that is symmetric about zero.

\subsection{Asymmetric quantum many-body scar}
\label{sec:dyn-fb}

Restoring the state-dependent FB interaction---the second term in Eq.~(\ref{eff_ham})---reverses this picture. The full model given by Hamiltonian~(\ref{eff_ham}) now hosts persistent oscillations for the polarized initial state $|\mathbf{0}\rangle$ rather than for $|\mathbb{Z}_4\rangle$. In particular, Fig.~\ref{fig4}(a) reveals coherent oscillations of a dimer stripe in the bulk of the chain, whereas no such oscillations occur for state $|\mathbb{Z}_4\rangle$, as shown in Fig.~\ref{fig4}(b). This nonthermal dynamics starting from the initial state $|\mathbf{0}\rangle$ is reminiscent of similar results in the PXP model when local detunings are not neglected~\cite{su_observation_2023}. The FB interaction tilts the scar structure of the kinetically constrained process away from state $|\mathbb{Z}_4\rangle$ and toward state $|\mathbf{0}\rangle$.



In Fig.~\ref{fig4}(c), we show the overlap between the polarized initial state $|\mathbf{0}\rangle$ and the eigenstates of $H$. As discussed in Sec.~\ref{sec:dyn-kin}, the kinetically constrained excitation process in Eq.~(\ref{eff_ham}) possesses a chiral symmetry $\mathcal{C} = \prod_{j=1}^{L/2-1} \sigma_{2j}^z$ and a spectrum symmetric about zero. The FB interaction breaks this symmetry, yielding an asymmetric spectrum. As seen in Fig.~\ref{fig4}(c), a strip of eigenstates away from the main cluster gains high overlaps with state $|\mathbf{0}\rangle$. Due to the high overlap, the dynamics largely retains the  memory of the initial state~\cite{Perciavalle2025}. These states possess low entanglement entropies [Fig.~\ref{fig4}(e)], consistent with nonthermalization of quantum many-body scars, and are responsible for the nonthermal dynamics. In contrast to the kinetic-only model described in Sec.~\ref{sec:dyn-kin}, where the high-overlap scars track state $|\mathbb{Z}_4\rangle$, the FB interaction shifts the scar tower toward state $|\mathbf{0}\rangle$.

The nearly equal spacing of these high-overlap eigenstates further supports their scar characteristics. As shown in Fig.~\ref{fig5}(a)-(b), the four eigenstates with the largest overlap with $|\mathbf{0}\rangle$ are embedded in, yet clearly distinguishable from, the connectivity graph of the constrained Hilbert space, and their average energy spacing remains approximately constant, $\Delta E\approx0.3$, weakly depending on chain length $L$. The nonergodic feature is also reflected in the long-range multi-body correlations, as shown in Fig.~\ref{fig5}(c) the connected correlation $C_{i,j}$ between site 11 and distant sites $j$ remains sizable for the initial state $|\mathbf{0}\rangle$, in stark contrast to the rapidly decaying correlations obtained from $|\mathbb{Z}_4\rangle$.
 In Fig.~\ref{fig6} we show the overlaps of state $|\mathbf{0}\rangle$ with the eigenstates for chain lengths from $L=20$ to $L=44$.  A strip of high overlaps is visible in every case, and the spacing among the four highest-overlap eigenstates remains approximately $0.3$. 
\begin{figure}[thbp]
    \centering
    \includegraphics[width=0.96\linewidth]{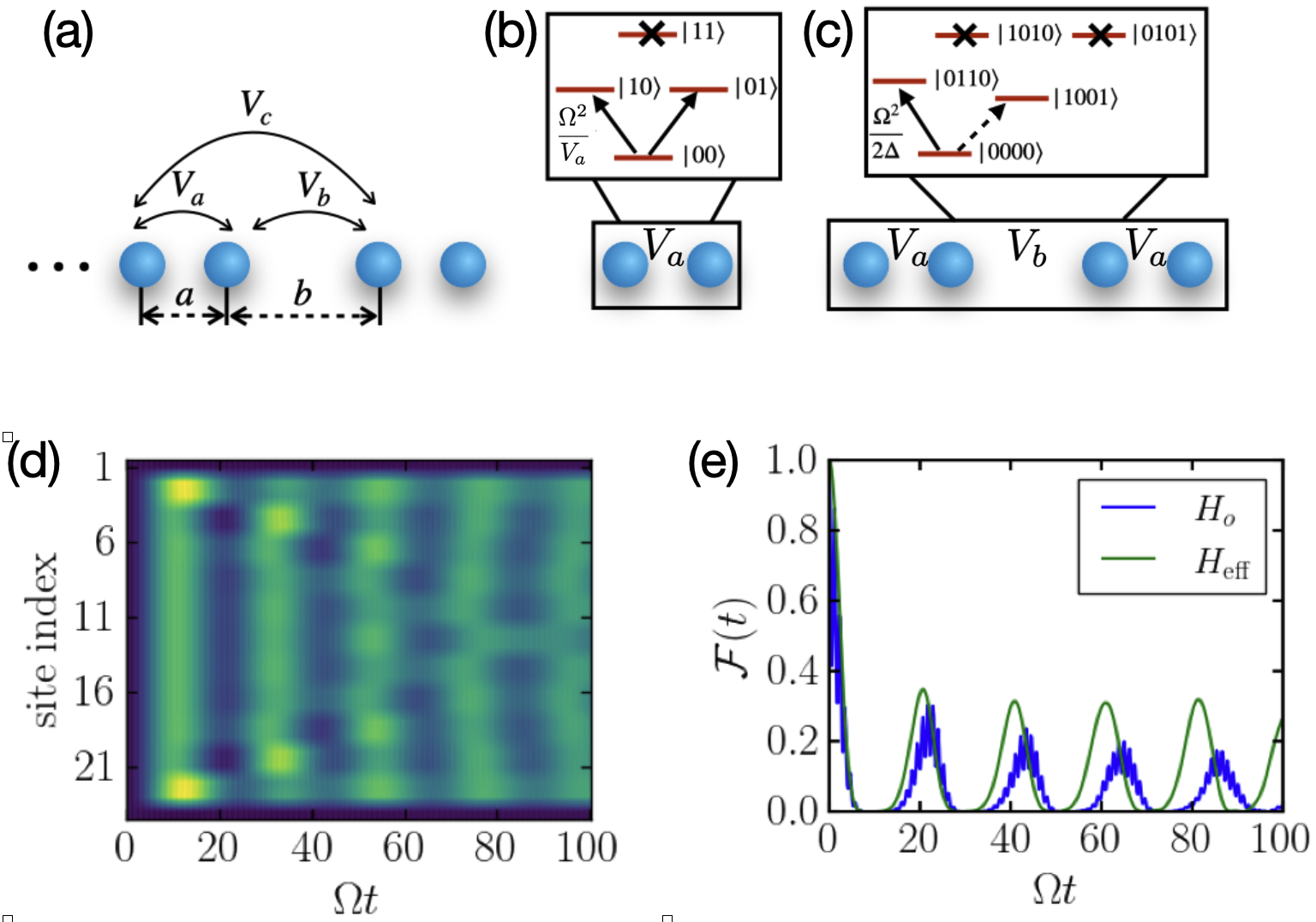}
    \caption{Realization of the kinetically constrained FB model with the Rydberg atom chain. (a) The interactions $V_a$ and $V_b$ in the shorter bond $a$ and longer bond $b$. The two bond lengths of the chain is $a$ and $b$, forming a Peierls array.  Even longer range interactions (e.g. $V_c$) are relatively weak, and do not affect the dynamics. (b) Rydberg blockade due to the interaction $V_a$. (c) Rydberg antiblockade between the atom pair in the middle of the tetramer. Due to the antiblockade of the pair interacting through $V_b$, transition $|0000\rangle\leftrightarrow |0110\rangle$ dominates in the dynamics, while transition $|0000\rangle\leftrightarrow |1001\rangle$ is largely suppressed.  (d) Occupation $\langle Q_j\rangle$ starting from state $|\mathbf{0}\rangle$. (e) Loschmidt echo of $|\mathbf{0}\rangle$ computed with $H_o$ and $H_{\mathrm{eff}}$. Both show oscillatory dynamics. In all panels, $L=24$, $\Omega/2\pi=1\,\mathrm{MHz}$, $\Delta/2\pi\approx 5.06\,\mathrm{MHz}$, $a=4\,\mu\mathrm{m}$, $b=8\,\mu\mathrm{m}$, and $C_6=2.655\times 10^6\mathrm{MHz}\cdot\mu\mathrm{m}^6$.}
    \label{fig7}
\end{figure}
\section{Implementation with a non-equidistant Rydberg-atom chain}
\label{sec:effective}

\subsection{Microscopic Hamiltonian and derivation}
\label{sec:derive}

We now show that the FB constrained Hamiltonian can be realized with a chain of Rydberg atoms. The system consists of a non-equidistant chain of $L$ Rydberg atoms, subjected to open boundary conditions (OBC), with alternating bond lengths $a$ and $b$, where $a<b$, as depicted in Fig.~\ref{fig7}(a). Atoms in the ground state $|0\rangle$ are resonantly coupled to the Rydberg state $|1\rangle$ by a laser with a Rabi frequency $\Omega$. Each site in the Rydberg state experiences local detuning $\Delta$. Pairs of atoms in the Rydberg state interact via van der Waals (vdW) potentials $V_{jk}=C_6/r_{jk}^6$, where $r_{jk}$ is the distance between sites $j$ and $k$, and $C_6$ is the dispersion coefficient. The many-body Hamiltonian ($\hbar\equiv1$) that governs the dynamics can then be written as $H_o=H_s+\sum_{j<k}V_{jk}Q_jQ_k$, where $H_s=\frac{\Omega}{2}\sum_{j}\sigma^x_j-\Delta\sum_{j}Q_j$ with ${\sigma}^x_j = |1\rangle_j\langle 0| + |0\rangle_j\langle 1|$ and $ {Q}_j = |1\rangle_j\langle 1|$.

Restricting to nearest-neighbor interactions, the Hamiltonian reduces to,
\begin{equation}
     {H}_o\approx H_s+ V_{a}\sum_{j=1}^{L/2} {Q}_{2j-1} {Q}_{2j}+ V_b\sum_{j=1}^{L/2-1} {Q}_{2j} {Q}_{2j+1}.
    \label{eq:Hatom}
\end{equation}
In our setting, we consider that the shorter bond $a$ generates strong interactions, $V_a \gg \Omega, \Delta $. As a result, the neighboring atom pair interacting through $V_a$ experiences the Rydberg blockade, as illustrated in Fig.~\ref{fig7}(b). We first simplify Eq.~(\ref{eq:Hatom}) using the rotating-wave approximation in order to removing the resulting double-excitation sector. Defining the unitary $ {U}=\exp{(-itV_{a}\sum_{j=1}^{L/2} {Q}_{2j-1} {Q}_{2j})}$, we transform the Hamiltonian according to $ {H}^{\prime}_o = {U} {H}_o {U}^{\dagger}$. Applying the blockade condition, and neglecting the fast oscillating terms of the pair excitation inhabited by interaction $V_a$, 
we obtain,
\begin{equation}
    \begin{split}
     \frac{{H}^{\prime}_o}{V_b}\approx &\underbrace{\sum_{j=1}^{L/2-1} {Q}_{2j} {Q}_{2j+1} - \frac{1}{2}\sum_{j=1}^{L} {Q}_{j}}_{ {H}_0} \\
     &+ \underbrace{\frac{\Omega}{2V_b}\sum_{j=1}^{L/2}\left( {\sigma}^x_{2j-1} {P}_{2j}+ {P}_{2j-1} {\sigma}_{2j}^{x}\right)}_{\lambda  {H}_1}.
    \end{split}
    \label{PX2}
\end{equation}
where we have divided the Hamiltonian by $V_b$. The reason of the dividing is that we can define small quantity $\lambda=\frac{\Omega}{2V_b}$, as $V_b\gg\Omega$. This allows us to distinguish the second line ($\lambda H_1$) on the right hand side of the equation as a perturbation. Note that $ {H}_0$ is purely diagonal while $ {H}_1$ is purely off-diagonal. Term $H_1$ describes essentially the Rydberg blockade process, where an atom can be excited to the Rydberg state only when its neighbor is in the electronic ground state. 



To this end we apply a Schrieffer--Wolff (SW) transformation. An anti-Hermitian generator $ {S}$ defines an effective Hamiltonian through $ {H}_{\text{eff}}=e^{\lambda{S}} {H}^{\prime}e^{-\lambda{S}}$. Expanding with the Baker--Campbell--Hausdorff formula yields,
\begin{equation}
    \begin{split}
     {H}_{\text{eff}}\approx& {H}_0 + \lambda \mathcal{P}\left([ {S}, {H}_0]+ {H}_1\right)\mathcal{P} \\
     &+ \lambda^2\mathcal{P}\left([ {S}, {H}_1]+\frac{1}{2!}[ {S},[ {S}, {H}_0]]\right)\mathcal{P} + \dots,
    \end{split}
\end{equation}
where $\mathcal{P}$ projects onto the desired subspace. $H_0$ and $H_1$ are defined in Eq.~(\ref{PX2}). Requiring $[ {S}, {H}_0]+ {H}_1=0$ determines $ {S}$,
\begin{widetext}
\begin{equation}
    \begin{split}
         {S} = 2i\Big[- ( {\sigma}_1^{y} {P}_2+ {P}_{L-1} {\sigma}_{L}^{y}) + \sum_{j=1}^{L/2-1}&( {P}_{2j-1} {\sigma}_{2j}^{y} {\sigma}_{2j+1}^{z} {P}_{2j+2}+  {P}_{2j-1} {\sigma}_{2j}^{y} {\sigma}_{2j+1}^{z} {Q}_{2j+2} \\
        &+  {Q}_{2j-1} {\sigma}_{2j}^{z} {\sigma}_{2j+1}^{y} {P}_{2j+2}+ {P}_{2j-1} {\sigma}_{2j}^{z} {\sigma}_{2j+1}^{y} {P}_{2j+2})\Big],
    \end{split}
    \label{s}
\end{equation}
\end{widetext}
in which the first two terms account for the open boundary conditions. 
Using the anti-blockade condition $  V_b = 2\Delta$ and projecting onto the subspace spanned by the null states of $ {H}_0$ (thereby breaking conservation of the total magnetization but preserving the inner pair magnetization) yields the effective Hamiltonian,
\begin{widetext}
\begin{equation}
    \begin{split}
        H_{\text{eff}} & \approx \frac{\Omega^2}{2\Delta}\sum_{j=1}^{L/2-1} {P}_{2j-1}( {\sigma}_{2j}^{+} {\sigma}_{2j+1}^{+} +  {\sigma}_{2j}^{-} {\sigma}_{2j+1}^{-}) {P}_{2j+2} \\
        &+ \underbrace{\frac{\Omega^2}{4\Delta}\sum_{j=1}^{L/2-1} ({P}_{2j-1} {\sigma}^{z}_{2j} {\sigma}_{2j+1}^{z} + {\sigma}^{z}_{2j} {\sigma}_{2j+1}^{z}{P}_{2j+2})- \frac{\Omega^2}{4\Delta}( {\sigma}_1^{z} {P}_2+ {P}_{L-1} {\sigma}_{L}^{z})}_{\mathcal{V}},
    \end{split}
    \label{eff_ham_full}
\end{equation}
\end{widetext}
where the first line on the right hand side of Eq.~(\ref{eff_ham_full}) is identical to the kinetic term in Hamiltonian~(\ref{eff_ham}). It becomes clear that we can relate the laser parameter with parameter $g=\Omega^2/2\Delta$. We now show that the second line, denoted by $\mathcal{V}$, gives the FB interaction in Hamiltonian~(\ref{eff_ham}).  


The matrix elements of part $\mathcal{V}$ in Eq.~(\ref{eff_ham_full}) are evaluated in the orthonormal ansatz basis $\{|\psi_m\rangle, m=0,2,4\dots L/2\}$ introduced in Sec.~\ref{sec:gs}. The diagonal matrix element $\langle\psi_m| {\mathcal{V}}|\psi_{m}\rangle$ is,
\begin{equation}
    \label{v_matrix}
    \langle\psi_m|\mathcal{V}|\psi_{m}\rangle = \frac{L}{8}\frac{\Omega^2}{\Delta}\mathbf{I}-\frac{\Omega^2}{8\Delta}(2m-L)
    =\frac{L}{4}\frac{\Omega^2}{\Delta}\mathbf{I}-\frac{\Omega^2}{4\Delta}m.
\end{equation}
In the space spanned by $|\psi_m\rangle$, it can be verified that the following term has the identical matrix element,
\begin{equation}
    \mathcal{V}' = \frac{L}{4}\frac{\Omega^2}{\Delta} {\mathbf{I}}-\frac{\Omega^2}{2\Delta}\sum_{j=1}^{L/2-1} {P}_{2j-1} {Q}_{2j} {Q}_{2j+1} {P}_{2j+2}. 
\end{equation}
In other words, both $\mathcal{V}$ and $\mathcal{V}'$ are identical in the space spanned by $|\psi_m\rangle$ except a constant term. As a result, we derive the effective Hamiltonian that is approximately given by,
\begin{equation}
    \begin{split}
    H_{\text{eff}} \approx &\frac{\Omega^2}{2\Delta}\sum_{j=1}^{L/2-1}\bigl[P_{2j-1}(\sigma_{2j}^{+}\sigma_{2j+1}^{+}+\text{h.c.})P_{2j+2} \\
    &- P_{2j-1}Q_{2j}Q_{2j+1}P_{2j+2}\bigr]+\frac{L}{4}\frac{\Omega^2}{\Delta} {\mathbf{I}}.
    \end{split}
    \label{eq:heff_equiv}
\end{equation}
Neglecting the constant term, and setting $g=\Omega^2/2\Delta$, we obtain the kinetically constrained FB interacting Hamiltonian~(\ref{eff_ham}). 

\subsection{Experimental parameters}
\label{sec:experiment}

We now show that the FB interaction Hamiltonian can be realized with realistic parameters. As an example, a non-equidistant Rydberg chain is realized using $^{87}$Rb atoms trapped in a one-dimensional tweezer array with programmable alternating spacings $a = 4~\mu\mathrm{m}$ and $b = 8~\mu\mathrm{m}$. Considering Rydberg state $|76D_{5/2}\rangle$, the vdW interaction is characterized by $C_6 = 2.655 \times 10^6~\mathrm{MHz}\cdot \mu\mathrm{m}^6$, yielding  nearest-neighbor interaction $V_a \approx 648~\mathrm{MHz}$ and interaction $V_b \approx 10.13~\mathrm{MHz}$. Global coherent driving is applied with a Rabi frequency $\Omega/2\pi = 1~\mathrm{MHz}$ and detuning $\Delta/2\pi \approx 5.06~\mathrm{MHz}$, satisfying the antiblockade condition $V_b = 2\Delta$ for atom pairs in a tetramer. 

\begin{figure}[thbp]
    \centering
    \includegraphics[width=0.96\linewidth]{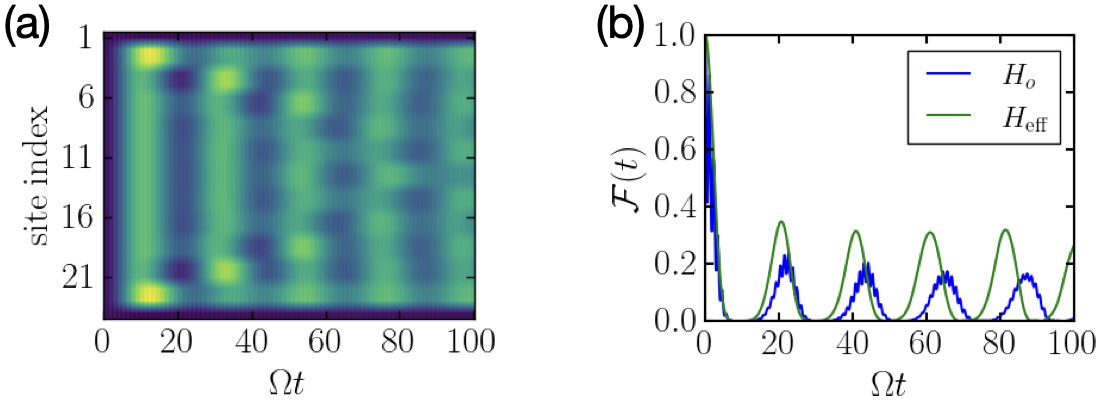}
    \caption{Same diagnostics as Fig.~\ref{fig7} with only nearest-neighbor couplings $V_a$ and $V_b$ retained in $H_o$. (a) Occupation $\langle Q_j\rangle$ starting from state $|\mathbf{0}\rangle$. (b) Loschmidt echo with initial state $|\mathbf{0}\rangle$ calculated with $H_o$ and $H_{\mathrm{eff}}$. Parameters as in Fig.~\ref{fig7}.}
    \label{fig8}
\end{figure}

With these parameters, we numerically calculate the dynamics using the original Hamiltonian $H_{o}$. In Fig.~\ref{fig7}(d), we show the Rydberg occupation dynamics under these parameters, which show coherent revivals from the polarized initial state $|\mathbf{0}\rangle$. In Fig.~\ref{fig7}(e), we compare the Loschmidt echo computed using the full Rydberg Hamiltonian $H_o$ and the effective model $H_{\text{eff}}$. The effective model qualitatively captures the nonergodic dynamics with initial state $|\mathbf{0}\rangle$. The same diagnostics with only the nearest-neighbor interaction $V_a$ and $V_b$ retained in $H_o$ are shown in Fig.~\ref{fig8}. The coherent revivals of state $|\mathbf{0}\rangle$ remain essentially unchanged relative to the Hamiltonian with the next-nearest neighbor interaction as depicted in Fig.~\ref{fig7}, confirming that the long-range tail of the van der Waals interaction does not affect the dynamics in this regime.

\section{Conclusion}
\label{sec:conclusion}

We have studied the many-body ground state and the weakly ergodicity-broken dynamics of a one-dimensional spin chain driven by the kinetically constrained FB interaction. In this model, a neighboring pair can be flipped from state $|0\rangle$ to $|1\rangle$ only when the two flanking sites are in state $|0\rangle$, after which the tetramer experiences a configuration-dependent clustering interaction. The two ingredients---a constrained flip and a state-dependent four-body potential---compete and produce a spectrum that is only weakly ergodic.

The ground state energy scales linearly with the chain length, $E_0\simeq -0.4148gL$ as $L\to\infty$, in agreement with exact diagonalization. The corresponding entanglement entropy oscillates with the subsystem size, reflecting strongly entangled bonds inside each active pair in a tetramer and weakly entangled bonds between tetramers.
With only the constrained flip, the model is similar to a clustered PXP Hamiltonian: it has a chiral symmetry, a spectrum symmetric about zero, and a tower of low-entanglement eigenstates with high overlap on state $|\mathbb{Z}_4\rangle$. Therefore, this state exhibits persistent bulk revivals, whereas state $|\mathbf{0}\rangle$ thermalizes. Restoring the FB interaction breaks the chiral symmetry, tilts the high-overlap strip away from state $|\mathbb{Z}_4\rangle$ and toward state $|\mathbf{0}\rangle$, and reverses the dynamics of the system. The scar energies remain nearly equally spaced, $\Delta E\approx 0.3$, largely independent of $L$, and initial state $|\mathbf{0}\rangle$ develops long-range connected correlations that are absent for state $|\mathbb{Z}_4\rangle$.



The proposed model is readily accessible on Rydberg atom simulation platforms. Multiple directions can be explored in this field. First, the long-range multi-body correlations that accompany the  revivals of state $|\mathbf{0}\rangle$ may be useful for quantum sensing, where a slowly decaying connected correlator can enhance the response to a weak perturbation. Second, the two terms in Eq.~(\ref{eff_ham}) share the same strength $g$, so they cannot be changed independently in the present regime. One could independently tune the strength for the two processes through, e.g. site-dependent detuning or laser driving. The resulting dynamics may even be exotic as the two processes can compete in the dynamics. Our study identifies the Rydberg antiblockade, used together with the blockade, as a route to constrained multi-body dynamics that violate the ETH weakly and can be probed with the Rydberg atom platform.

\begin{acknowledgments}
TY acknowledges discussions with Stephen Powell, Arash Jafarizadeh and Guoxian Su. We acknowledge support from the EPSRC through Grant Nos.~EP/W015641/1 and EP/W524402/1, and the use of the University of Nottingham's Ada HPC service.
\end{acknowledgments}

\textit{AI usage.} The authors used \emph{Opencode (Big Pickle)} for grammar and spell-checking of the manuscript. All suggestions were reviewed and approved by the authors.

\bibliography{references}

\end{document}